\documentclass[aps,prl,twocolumn,superscriptaddress]{revtex4-2}
\usepackage{graphicx}
\usepackage{epstopdf}
\usepackage{bm}
\usepackage{latexsym} 
\usepackage{braket}
\usepackage[colorlinks,linkcolor=blue,citecolor=blue,urlcolor=blue]{hyperref}
\usepackage{amssymb}
\usepackage{amsmath}
\usepackage{mathtools}
\usepackage{color}
\usepackage{import}
\usepackage[caption = false]{subfig}
\usepackage{import, xcolor}
\usepackage[retainorgcmds]{IEEEtrantools}
\usepackage{bm}
\usepackage[none]{hyphenat}
\usepackage{textcomp} 

\newcommand{\beq}{\begin{equation}}
\newcommand{\eeq}{\end{equation}}

\newcommand{\un}[1]{\mathrm{\:#1}} 
\renewcommand{\vec}[1]{\boldsymbol{#1}} 
\newcommand{\ii}{i}
\newcommand{\ee}{\text{e}}

\begin{document}

\title{Suppression of Nonlinear Optical Rogue Wave Formation Using Polarization-Structured Beams}
\author{A. Nicholas Black}
\affiliation{Department of Physics and Astronomy, University of Rochester, Rochester, NY, USA 14627}
\author{Saumya Choudhary}
\affiliation{Institute of Optics, University of Rochester, Rochester, NY, USA 14627}
\author{E. Samuel Arroyo-Rivera}
\affiliation{Department of Physics and Astronomy, University of Rochester, Rochester, NY, USA 14627}
\author{Hayden Woodworth}
\affiliation{Institute of Optics, University of Rochester, Rochester, NY, USA 14627}
\author{Robert W. Boyd}
\affiliation{Institute of Optics, University of Rochester, Rochester, NY, USA 14627}
\affiliation{Department of Physics, University of Ottawa, Ottawa, ON, Canada K1N 6N5}

\begin{abstract}
A nonlinear self-focusing material can amplify random small-amplitude phase modulations present in an optical beam, leading to the formation of amplitude singularities commonly referred to as optical caustics.  By imposing polarization structuring on the beam, we demonstrate the suppression of amplitude singularities caused by nonlinear self-phase modulation.  Our results are the first to indicate that polarization-structured beams can suppress nonlinear caustic formation in a saturable self-focusing medium and add to the growing understanding of catastrophic self-focusing effects in beams containing polarization structure.
\end{abstract}
\maketitle
Physical systems governed by wave mechanics are capable of evolving into configurations that concentrate energy into small regions of space.  In optics, the concentration of nearly-parallel rays corresponding to different wavefronts into a small area is known as caustic formation~\cite{BerryPiO80,Born99}.  A familiar example of this behavior is the pattern of light formed on the bottom of a swimming pool caused by refraction from small waves on the water surface.  Caustics have also been observed in the phase space trajectories of a driven two-level atomic system~\cite{NaghilooPRA17}.  Caustic formation is a fundamentally linear phenomenon arising from the diffraction of light fields containing random phase perturbations~\cite{BerryPiO80,Born99}.  However, nonlinear self-phase modulation can enhance the formation of caustics from small initial phase perturbations~\cite{SafariPRL17,PierangeliPRL15}.  In addition to caustic enhancement, self-action effects are responsible for the breakup of laser beams into small-scale filaments~\cite{Boyd2020}, the catastrophic collapse of beams into point-like regions of space, and the formation of nondiffracting beams known as spatial solitons~\cite{ChiaoPRL64,ChenRepProgPhys12}.  The dynamical equations governing the motion of ocean waves also contain a self-phase modulation term that leads to the formation of rogue waves~\cite{ChabchoubPRL14}. 

Optical communications~\cite{WrightNatPhot16}, remote sensing, and lightning strike control~\cite{KoopmanJApplPhys71, ZhaoJQE95,KasparianSci03, ProduitEPJApplPhys20} are a few technologies that rely upon a careful understanding of the interplay between linear and nonlinear propagation effects.  Encoding information in the orbital angular momentum (OAM) of light is a promising way to increase the information capacity of optical communication channels~\cite{GibsonOptExp04,BozinovicSci13,YanNatComm14,ZhouNatComm21}.  However, the amplification of azimuthal modulation instabilities cause OAM beams to break up during nonlinear propagation~\cite{FirthPRL97,BigelowPRL04,VuongPRL06}.  Beams carrying a space-varying polarization have been suggested as an alternative encoding scheme~\cite{LarocqueNatComm20} that follows an algebra similar to the Poincar\'e sphere formalism for plane wave polarization~\cite{MilionePRL11}.  Such beams are typically referred to as polarization-structured beams.  As an added benefit, certain polarization-structured beams form the normal-mode basis of optical fiber waveguides~\cite{RamachandranOptLett09,MilionePRL11,SitOptLett18}.  

Polarization-structured beams are solutions to the vector paraxial wave equation and can be categorized by the number of polarization states represented in their cross section.  Radially and azimuthally polarized beams, typically called \textit{vector vortex} beams~\cite{ZhanAOP09}, trace a path on the Poincar\'e sphere, parameterized by the azimuthal angle in the beam's cross-section.  Another class of beams, known as \textit{full-Poincar\'e} beams, sweep out the entire surface area of the Poincar\'e sphere, parameterized by both the radial and azimuthal location in the beam~\cite{BeckleyOptExp10}.  Examples of full-Poincar\'e beams include \textit{lemon}, \textit{star}, and \textit{monstar} topologies~\cite{Gbur17}.  There is a final class of beams involving partially polarized light, referred to here as \textit{volumetrically-full Poincar\'e} beams, that sweep out the entire volume of the Poincar\'e sphere, parameterized by radial, azimuthal, and axial position in the beam~\cite{BeckleyOptExp12}. 

Recent theoretical~\cite{SaPRA19} and experimental~\cite{BouchardPRL16} results have shown that vector vortex and full-Poincar\'e beams are less prone to self-focusing and nonlinear beam breakup, with full Poincar\'e beams being the most resistant.  Full-Poincar\'e beams are also less prone to linear beam breakup caused by atmospheric turbulence~\cite{GuOptLett09}.  Conversely, certain polarization-structured beams formed from Hermite-Gauss modes are more susceptible to nonlinear collapse but in a predictable way that is stable against random phase modulations~\cite{LiSciRep12,HuOptLett21}.  These beams, sometimes referred to as hybrid vector beams, are the polarization-structured analog of necklace beams~\cite{SoljacicPRL98}, which are known to have stable propagation in nonlinear self-focusing media.

In this Letter, we show through both experiment and simulation that full-Poincar\'e beams are less likely to develop caustics upon nonlinear propagation compared to a uniformly polarized Gaussian beam and to a uniformly polarized beam with the same intensity structure as full-Poincar\'e beams.  We study this suppression of nonlinear caustic formation in a saturable, nonlinear self-focusing medium.  These findings add to the growing understanding of rogue phenomena and are the first to address nonlinear caustic formation in polarization-structured beams.

\section{Background}
Fully coherent polarization-structured beams can be decomposed into a superposition of orthogonally-polarized transverse spatial modes,
\begin{equation}
    \vec{\mathrm{E}}\mathrm{(\rho,\phi, z,t)} = \left[E_{a}(\rho,\phi,z)\vec{\mathrm{e}}_a + E_{b}(\rho,\phi,z)\vec{\mathrm{e}}_b\right]\ee^{-\ii\omega t},
    \label{eq:genpolstruct}
\end{equation}
where $\vec{\mathrm{e}}_a$ and $\vec{\mathrm{e}}_b$ are (generally complex) orthogonal unit vectors.  The inseparability of polarization and spatial mode in Eq.~(\ref{eq:genpolstruct}) has been the subject of investigations and debate about its connection with measures of quantum entanglement~\cite{EberlyPhysScrp16,KarimiSci15}, though it describes a purely classical beam.  Because the orthogonally polarized modes in Eq.~(\ref{eq:genpolstruct}) do not interfere with each other, some have suggested this as the reason for polarization-structured beams' stability against beam breakup~\cite{GuOptLett09,Khare20}.  For lemon and star beams, $E_a$ and $E_b$ are the Laguerre-Gauss modes $\mathrm{LG_{0,0}(\rho,\phi,z)}$ and $\mathrm{LG_{0,\pm1}(\rho,\phi,z)}$, respectively.  The first subscript in $\mathrm{LG}_{p,l}$ denotes the radial index and the second denotes the azimuthal index.  Lemon and star beams are differentiated by the azimuthal index of the mode $E_b$, with a lemon beam having $l = +1$ and a star beam having $l = -1$.

The nonlinear propagation of beams described by Eq.~(\ref{eq:genpolstruct}) can be modeled using coupled-mode Helmholtz equations~\cite{Agrawal19,Boyd2020,FeitJOSAB88},
\begin{equation}
    \begin{aligned}
        \nabla^{2}E_{a} &= -k_{0}^{2}\left(1 + \chi_{a}\right)E_{a} \\
        \nabla^{2}E_{b} &= -k_{0}^{2}\left(1 + \chi_{b}\right)E_{b},
    \end{aligned}
    \label{eq:NLSE}
\end{equation}
where
\begin{equation}
    \chi_{i} = \chi^{(1)} + \frac{8 n_0 \epsilon_0 c n_2}{3}\frac{|E_i|^2 + \mu|E_j|^2}{1 + \sigma\left(|E_i|^2 + \mu|E_j|^2\right)},
    \label{eq:susceptibility}
\end{equation}
for $i \neq j$. In Eqs.~(\ref{eq:NLSE}) and ~(\ref{eq:susceptibility}), $k_0$ is the free-space wave number, $\chi^{(1)}$ is the linear susceptibility, $n_0$ is the linear refractive index, and $n_2$ is the intensity-dependent refractive index.  

\begin{figure}
    \centering
    \includegraphics[width = \columnwidth]{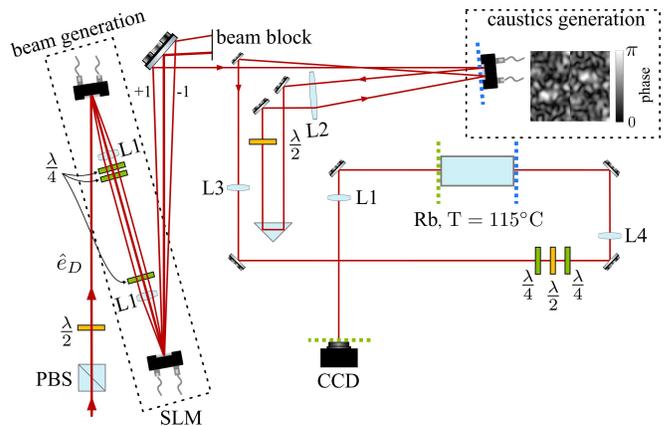}
    \caption{The experimental setup for measuring spatially resolved intensity statistics.  A diagonally polarized narrow-linewidth laser beam ($+0.6\un{GHz}$ above the $^{87}\mathrm{Rb}$ $\mathrm{D}_2$ $\mathrm{F} = 1 \rightarrow \mathrm{F'} = 2$ transition) enters a system of two spatial light modulators (SLM) capable of generating any fully coherent polarization-structured beam (beam generation).  Each SLM acts upon a different orthogonal polarization component of the beam, and the face of the first SLM is imaged onto the face of the second using a 4$f$ system.  A third SLM divided into two regions, imparts the same spatially-random phase to each polarization component of the beam. The face of the third SLM (dotted blue line) is imaged onto the entrance facet of a Rb cell using a Keplerian telescope with a magnification of $-3/4$ (L3 and L4).  The polarization is transformed to the circular basis using a series of half- and quarter-wave plates ($\lambda/2$ and $\lambda/4$, respectively).  The output facet of the Rb cell (dotted green line) is imaged onto a CCD camera (CCD) to collect pixel intensity statistics.  The lens focal lengths are $f = 20\un{cm}$ (L1), $f = 30\un{cm}$ (L2), $f = 1\un{m}$ (L3), $f = 75\un{cm}$ (L4); polarizing beamsplitter (PBS).}
    \label{fig:expsetup}
\end{figure}

Equation~(\ref{eq:susceptibility}) is a commonly used phenomenological model of cross-phase modulation in a saturable medium~\cite{BigelowPRE02,SalgueiroPRE04,LiSciRep12,BouchardPRL16}.  The cross-coupling coefficient, $\mu$, varies depending upon the nonlinear material under consideration. For atomic vapor nonlinearity, the specific atomic level system under consideration dictates the value of $\mu$~\cite{ChengPCCP19} and can even lead to the arrest of self-focusing under conditions of coherent population trapping~\cite{JainPRL95}. The saturation coefficient, $\sigma$, is proportional to the inverse of the saturation intensity.  The experimental configuration shown in Fig.~\ref{fig:expsetup} is well described by $n_0 = 1 - 6\times10^{-5}$, $n_2 = 1.5\times10^{-10}\ \un{m^2/W}$~\cite{McCormickPRA04}, $\mu = 0.3$, and $\sigma = 3.9\times10^{-9}\ \un{m^2/V^2}$.  All values except $\mu$ and $\sigma$ were obtained by considering the particular two-level transition described in Fig.~\ref{fig:expsetup}~\cite{Boyd2020,Steck21}.  $\mu$ and $\sigma$ were obtained by qualitatively matching the simulation of Eq.~\ref{eq:NLSE} with experiment results.

From the right-hand-side of Eqs.~(\ref{eq:NLSE}) and~(\ref{eq:susceptibility}) it is clear that the refractive index experienced by one mode is influenced by the intensity profile of the other mode.  The resulting cross-coupling behavior~\cite{VudyasetuPRL09} leads to the modification of the self-focusing distance of the composite beam, as mentioned in Refs.~\cite{LiSciRep12,SaPRA19,HuOptLett21}.  In the absence of cross-coupling ($\mu = 0$), the intensity-dependent susceptibility on the right-hand-side of Eqs.~(\ref{eq:NLSE}) leads to the enhancement of caustic formation through self-focusing~\cite{PierangeliPRL15,SafariPRL17}.  This effect is the result of small phase perturbations being amplified through a four-wave mixing process in the nonlinear medium~\cite{Boyd2020}.
\begin{figure*}
    \centering
    \includegraphics[width = \textwidth]{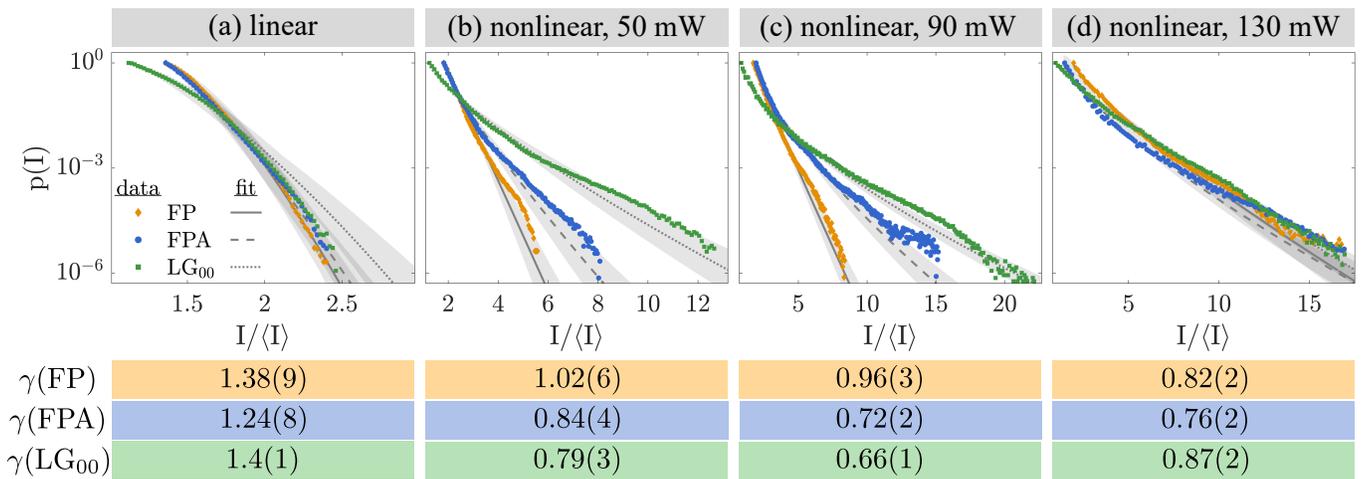}
    \caption{The experimentally obtained intensity statistics of FP (orange diamonds), FPA (blue circles), and $\mathrm{LG_{0,0}}$ (green squares) beams after (a) linear and (b,c,d) nonlinear propagation.  The solid, dashed, and dotted lines are Eq.~(\ref{eq:statistics}) fitted to the FP, FPA, and $\mathrm{LG_{0,0}}$ data, respectively.  The value of $\gamma$ obtained from fitting Eq.~(\ref{eq:statistics}) to each dataset is shown at the bottom of the figure.  The shaded area indicates the one standard deviation uncertainty in the fits.  Under linear propagation, all beams have very similar intensity statistics that display no caustic formation.  Under nonlinear propagation, the uniformly polarized $\mathrm{LG_{0,0}}$ and FPA beams begin to display caustic formation that increases with increasing beam power (b and c).  For the same beam powers, the polarization-structured FP beam maintains exponential intensity statistics with no caustics present.  The suppression of caustics afforded by the polarization structure of the beam is no longer present at a beam power of $130\un{mW}$.}
    \label{fig:tailvspower}
\end{figure*}

The spatially-resolved intensity statistics of beams undergoing breakup can be modeled by the following probability density function~\cite{PierangeliPRL15},
\begin{equation}
    p(I) = N\ee^{-\zeta\left(\frac{I}{\langle I \rangle}\right)^{\gamma}},
    \label{eq:statistics}
\end{equation}
where $N$ is a normalization coefficient, $\zeta$ describes the width of the distribution, $\gamma$ describes the tails of the distribution, and $I$ is the intensity at each transverse location in the beam.  Brackets denote an average over the transverse spatial coordinates.  Fully-developed speckle patterns follow exponential intensity statistics, corresponding to $\gamma = 1$~\cite{Goodman20}.  Upon the development of caustics, intrabeam intensities will begin to follow a long-tailed distribution, characterized by $0 < \gamma < 1$ ~\cite{PierangeliPRL15}.  A long-tailed intensity distribution is an indicator of rogue high intensity peaks within the beam, in analogy with rogue ocean waves.

\section{Experiment}
The experimental setup for generating full-Poincar\'e beams and measuring caustic formation is shown in Fig.~\ref{fig:expsetup}.  To generate full-Poincar\'e beams, a narrow linewidth $(\sim\! 200\un{kHz})$ diagonally polarized Gaussian beam with a diameter of $\sim\! 5\un{mm}$ enters a system of two spatial light modulators (SLM).  The first SLM is programmed with a blazed computer-generated hologram (CGH) that encodes an $\mathrm{LG_{0,1}}$ onto a carrier spatial frequency~\cite{ArrizonJOSAA07} for the horizontally polarized portion of the beam only.  The face of the first SLM is imaged onto the second SLM using a $4f$ system, and the polarization of the beam is rotated by $90^\circ$ so that the second SLM acts only on the portion of the beam that did not interact with the first SLM.  The second SLM then uses a CHG to encode an $\mathrm{LG_{0,0}}$ onto a carrier spatial frequency that exactly overlaps with the $\mathrm{LG_{0, 1}}$ created on the first SLM.  The diameter of the generated beam $(\sim\! 1\un{mm})$ is approximately five times smaller than the diameter of the input Gaussian beam, resulting in minimal influence of the underlying Gaussian structure on the generated beam.  This scheme has the advantage of being configured to generate any fully coherent polarization structured beam within the spatial bandwidth of the SLMs.  

The generated beam then travels to a third SLM that imprints a random phase mask on both polarization components of the beam.  To do this, the SLM is divided into two regions, and one region is imaged onto the other using a $4f$ system containing a $90^\circ$ polarization rotation.  The correlation length of random phase features in the mask is $450\un{\mu m}$, and the maximum phase shift in the mask is $\pi$ radians.  The face of the third SLM is imaged onto the input facet of a Rb vapor cell using a Keplerian telescope with a magnification of $-3/4$, and the polarization of the beam is transformed to the circular basis.  At the input of the vapor cell, the full-Poincar\'e beam has a lemon topology with an overall random phase,
\begin{equation}
    \vec{\mathrm{E}}(\rho, \phi, z) = \ee^{i\phi_{\mathrm{rand}}(\rho,\phi,z)}\left[\mathrm{LG_{0,0}}\vec{\mathrm{e}}_\mathrm{L} + \mathrm{LG_{0,1}}\vec{\mathrm{e}}_\mathrm{R}\right]
    \label{eq:beam}
\end{equation}

The vapor cell contains natural abundance Rb and is heated to $115\un{^\circ C}$ to achieve a high number density of Rb atoms $\left(\sim\! 10^{19} \un{atoms/m^{3}}\right)$ in the cell.  The laser is blue-detuned to $+0.6\un{GHz}$ above the $^{87}\mathrm{Rb}$ $\mathrm{D}_2$ $\mathrm{F} = 1 \rightarrow \mathrm{F'} = 2$ transition and experiences a self-focusing nonlinearity.  The field at the output facet of the Rb cell is then imaged onto a camera to collect pixel intensity statistics.

\section{Results}
Figure~\ref{fig:tailvspower} shows intensity statistics collected for three different beams: a lemon beam (FP), a uniformly polarized beam with the same intensity structure as a lemon beam (FPA), and a uniformly polarized $\mathrm{LG_{0,0}}$ beam with the same beam waist as the $\mathrm{LG_{0,0}}$ component of the FP beam.
Each histogram is comprised of pixel intensities from the imaging camera for 500 trials with different random phase masks.  The frames from the camera are truncated to include only pixels within a region that contains nonzero intensity when all frames from the 500 trial collection are averaged together.  The frame sizes for the FP and FPA trials are similar, but the frame sizes are smaller for the $\mathrm{LG_{0,0}}$ trials--as expected for the smaller $\mathrm{LG_{0,0}}$ beam.  The pixel intensities are divided by the average intensity observed in all trials.  Equation~(\ref{eq:statistics}) is fit to the tails of the histograms using maximum likelihood estimation to measure the ``tailiness" of the intensity distribution.  Uncertainties in the fits were obtained through Monte Carlo simulation.  

When the beam power is low ($\sim 4\un{mW}$) and the Rb vapor is at room temperature, the beams propagate linearly through the cell, Fig.~\ref{fig:tailvspower}(a).  Under linear propagation, the FP, FPA, and $\mathrm{LG_{0,0}}$ behave almost identically with no caustic formation present, as indicated by $\gamma > 1$.  In fact, the beams do not even display speckle pattern statistics under these conditions.  In Fig.~\ref{fig:tailvspower}(b) the cell temperature is increased to $115\un{^\circ C}$ and the beam powers are increased to $50\un{mW}$. The $\mathrm{LG_{0,0}}$ and FPA beams begin to develop long-tailed intensity statistics, $\gamma < 1$, indicating the presence of caustics.  Under these same conditions, the FP beam displays speckle-pattern statistics.  Thus, the tendency of the beam to display nonlinear caustics is seen to be suppressed through use of a polarization structured beam. As the beam powers are increased to $90\un{mW}$, Fig.~\ref{fig:tailvspower}(c), the intensity histograms for the $\mathrm{LG_{0,0}}$ and FPA beams develop longer tails while the FP beam maintains Gaussian amplitude statistics.  At the maximum achievable beam power of our system, $130\un{mW}$, all beams display similar long-tailed intensity statistics, Fig.~\ref{fig:tailvspower}(d).  

The suppression of caustic enhancement in the polarization-structured beam can be attributed to two primary effects.  The first is that FP beams can be treated as a mutually incoherent superposition of an $\mathrm{LG_{0,0}}$ and an $\mathrm{LG_{0,\pm 1}}$ beam.  This leads to the decrease in linear beam breakup from phase perturbations~\cite{GuOptLett09} through an effect known as complementary diffraction, as described in Ref.~\cite{Khare20}.  The second effect contributing to the suppression of nonlinear caustic formation is the cross-phase modulation between the two modes comprising the FP beam.  Mutual interaction can stabilize the beam under nonlinear propagation~\cite{BouchardPRL16,BigelowPRE02,SaPRA19}.  The suppression of caustic enhancement afforded by polarization structure does not persist at higher powers, as indicated by the results of Fig.~\ref{fig:tailvspower}(d).  Furthermore, caustic enhancement also appears to saturate at higher powers because $\gamma$ has increased for all but the FP beam at $130\un{mW}$.  

\begin{figure}
    \centering
    \includegraphics[width = \columnwidth]{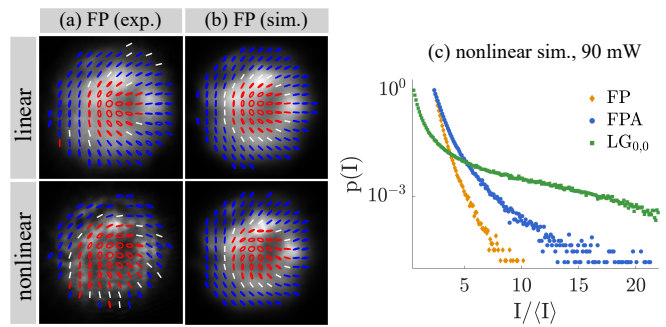}
    \caption{Comparison of experimental and simulation results for linear and nonlinear propagation at a beam power of $90\ \un{mW}$.  The polarization handedness in (a) and (b) is indicated by color, where red, blue, and white indicate left-circular, right-circular, and linear polarization respectively.  The same random phase mask is used in (a) and (b).  The intensity structure does not change dramatically after linear propagation because the maximum phase of the phase mask is many times smaller than the maximum phase at which caustics usually develop ($\sim\!\!8\pi\un{rad.}$) (top). However, the intensity structure changes dramatically upon nonlinear propagation over the same distance (bottom).  Nonetheless, the polarization structure remains similar to the linear result. The polarimetry simulation (b) shows good qualitative agreement with the experimental results (a).  (c)  In the numerical simulation of nonlinear propagation with $500$ different random phase masks, the FP beam has shorter-tailed statistics than either the FPA or $\mathrm{LG_{0,0}}$ beams, in agreement with experiment (Fig.~\ref{fig:tailvspower}).  The random phase masks used in obtaining (c) have the same parameters as those used in experiment.}
    \label{fig:causticssimulation}
\end{figure}
Figure~\ref{fig:causticssimulation} compares the experimentally-obtained polarization and intensity structure of the FP beam (Fig.~\ref{fig:causticssimulation}(a)) to numerical simulation (Fig.~\ref{fig:causticssimulation}(b)) for both linear and nonlinear propagation after the implementation of a random phase mask.  The simulation was performed by numerically solving Eqs.~(\ref{eq:NLSE}) using a split-step Fourier method that accounts for nonparaxiality~\cite{FeitJOSAB88} with a  beam power of $90\un{mW}$. The simulation results display good qualitative agreement with experiment.  Under linear propagation, the intensity structure does not change dramatically because the maximum of the random phase mask ($\pi \un{rad.}$) is small compared with that which typically leads to caustics ($\sim\!8\pi\un{rad.}$)\cite{SafariPRL17}. Remarkably, after nonlinear propagation, the lemon polarization topology of the FP beam changes very little, despite the dramatic change in its intensity structure.  This is likely due to the fact that the fields in each circular polarization component of the FP beam experience similar nonlinear phase shifts due to the cross-phase terms in Eqs.~\ref{eq:NLSE}.  If the coupling coefficient, $\mu$, were equal to unity, the polarization structure would not change at all because both circular polarization components would experience the exact same nonlinear phase.  That is, there would be no nonlinear birefringence~\cite{Agrawal19}.  In this case, the polarization structure would be the same for both linear and nonlinear propagation results.

In Fig.~\ref{fig:causticssimulation}(c) we simulate the intensity statistics of FP, FPA, and $\mathrm{LG_{0,0}}$ beam after nonlinear propagation for beam powers of $90\ \un{mW}$.  The FP beam is the least likely to develop rogue intensity peaks and the $\mathrm{LG_{0,0}}$ is the most likely to develop rogue intensity peaks, in agreement with experiment.  Compared with experiment, the FP beam has slightly longer-tailed statistics in simulation.  This is likely due to small differences in the intensity structure of the beam in experiment and simulation.  Nonetheless, the model of Eqs.~(\ref{eq:NLSE}) and~(\ref{eq:susceptibility}) describe the experimental results quite well.
\section{Conclusion}
We have shown that a full-Poincar\'e lemon beam is less susceptible to developing caustics upon propagation through a saturable, nonlinear self-focusing medium than either a uniformly polarized beam with the same intensity structure or a uniformly polarized $\mathrm{LG_{0,0}}$ beam with the same waist as the $\mathrm{LG_{0,0}}$ component of the lemon beam.  We simulate our experiment by numerically solving coupled-mode Helmholtz equations for a beam propagating through a saturable self-focusing medium and obtain good agreement with experiment. Our results add to the growing understanding of rogue behavior~\cite{PierangeliPRL15, SafariPRL17}, and they bear upon the use of polarization-structured beams to control nonlinear self-focusing processes in remote sensing~\cite{KoopmanJApplPhys71, KasparianSci03, ProduitEPJApplPhys20}, optical communications~\cite{SitOptLett18}, and laser engineering~\cite{JordanOptLett94}.  The extent to which polarization structure is maintained during nonlinear self-focusing warrants further study.  Such investigations would extend the field of singular optics into the nonlinear domain~\cite{SoskinJEOSA98,Gbur17}, potentially revealing topologically protected quantities that could be used for information transfer.

\begin{acknowledgements}
We would like to acknowledge fruitful discussions with Giulia Marcucci and Jerry Kuper.  We thank Aku Antikainen for proofreading our simulation code and correcting a minor bug. This work was supported by the US Office of Naval Research under awards N00014-19-1-
2247 and MURI N00014-20-1-2558.
\end{acknowledgements}

\bibliography{SuppressionofNonlinearCausticsRef}
\end{document}